\documentclass[prodmode,acmtecs]{acmsmall}

\usepackage{footmisc}
\usepackage{xcolor}
\usepackage{enumitem}
\usepackage{extarrows}
\usepackage{amsmath}
\usepackage{amssymb}
\usepackage{amsfonts}
\usepackage{mathrsfs}
\usepackage{graphicx}
\usepackage[ruled]{algorithm2e}
\usepackage{thm-restate}
\usepackage[colorlinks = true]{hyperref}
\hypersetup{
    linkcolor=black,
    citecolor=black,
    urlcolor=blue}

\usepackage{mathtools}
\usepackage{caption}

\usepackage{tikz}
\usetikzlibrary{shapes,snakes,backgrounds,arrows,patterns,positioning,arrows.meta,calc,automata}

\allowdisplaybreaks

\newtheorem{rmk}[theorem]{Remark}

\newcommand{\red}{\textcolor{red}}
\newcommand{\purple}{\textcolor{purple}}
\newcommand{\blue}{\textcolor{blue}}

\newcommand{\Z}{\mathbb{Z}}
\newcommand{\N}{\mathbb{N}}
\newcommand{\Q}{\mathbb{Q}}

\newcommand{\K}{\mathbb{K}}

\newcommand{\UT}{\mathsf{UT}}
\newcommand{\SL}{\mathsf{SL}}
\newcommand{\GL}{\mathsf{GL}}

\newcommand{\T}{\mathsf{T}}

\newcommand{\HH}{\operatorname{H}}

\newcommand{\mL}{\mathcal{L}}
\newcommand{\mA}{\mathcal{A}}

\newcommand{\ba}{\boldsymbol{a}}

\newcommand{\bzer}{\boldsymbol{0}}

\newcommand{\SA}{\mathsf{SA}(2, \mathbb{Z})}

\newcommand{\mG}{\mathcal{G}}
\newcommand{\sgmG}{\langle \mathcal{G} \rangle}
\newcommand{\gmG}{\langle \mathcal{G} \rangle_{grp}}
\newcommand{\mH}{\mathcal{H}}

\newcounter{ProblemCounter}

\begin{document}

\title{Recent advances in algorithmic problems for semigroups}

\author{Ruiwen Dong
\affil{Department of Computer Science, University of Oxford}
}



\begin{abstract}
In this article we survey recent progress in the algorithmic theory of matrix semigroups.
The main objective in this area of study is to construct algorithms that decide various properties of finitely generated subsemigroups of an infinite group $G$, often represented as a matrix group.
Such problems might not be decidable in general.
In fact, they gave rise to some of the earliest undecidability results in algorithmic theory.
However, the situation changes when the group $G$ satisfies additional constraints.
In this survey, we give an overview of the decidability and the complexity of several algorithmic problems in the cases where $G$ is a low-dimensional matrix group, or a group with additional structures such as commutativity, nilpotency and solvability.
\end{abstract}

\maketitle

\section{Introduction}\label{sec:intro}
It has been known since the work of Church and Turing from the 1930s that certain decision problems in mathematical logic do not admit algorithmic solutions.
For a long period of time, all examples of undecidable problems were derived directly from mathematical logic or the theory of computing, and the notion of undecidability seemed intangible for most mathematicians.
It was not until the late 1940s that the Soviet mathematician Andrey Markov produced a concrete undecidable problem using linear algebra.
In his seminal work \emph{``On certain insoluble problems concerning matrices''}~\cite{markov1947certain}, Markov studied the following decision problem.
Its input is a finite set of square matrices $\mG = \{A_1, \ldots, A_K\}$ and a matrix $T$, and the problem is whether or not there exist an integer $p \geq 1$ and a sequence $A_{i_1}, \ldots, A_{i_p}$ of matrices in $\mG$ such that $T = A_{i_1} A_{i_2} \cdots A_{i_p}$.
Markov showed this problem to be undecidable for integer matrices of dimension at least six, thus marking the first undecidability result obtained outside of mathematical logic and the theory of computing.

Markov's work falls into the area of computational group theory, which is one of the oldest and most well-developed parts of computational algebra. 
The ``official'' start of computational group theory dates back to 1911, when Max Dehn formulated three basic problems that would become its foundation.
Given a finite presentation of a group $G$, it is asked whether there are algorithms that solve the \emph{Word Problem} (whether an element is the neutral element), the \emph{Conjugacy Problem} (whether two elements are conjugate in $G$), and the \emph{Isomorphism Problem} (whether $G$ is isomorphic to another finitely presented group).
It was not until the 1950s that the three problem were shown to be undecidable in general groups~\cite{novikov1955algorithmic,adyan1955algorithmic}.

Using the language of computational group theory, Markov's problem can be reformulated as deciding \emph{Semigroup Membership} in a matrix (semi)group.
For a finite subset $\mG$ of a group $G$, denote by $\sgmG$ the subsemigroup generated by $\mG$.
Then the Semigroup Membership problem can be formulated as follows.
\begin{enumerate}[noitemsep, label = (\roman*)]
    \item \textit{(Semigroup Membership)} given a finite set $\mG$ and an element $T$ in $G$, decide whether $T \in \sgmG$.
    \setcounter{ProblemCounter}{\value{enumi}}
\end{enumerate}
Markov's undecidability result as well as the subsequent undecidability results for Dehn's problems generated a surge of research interest in computational group theory.
In the 1960s, Mikhailova introduced the group version of Semigroup Membership.
For a finite subset $\mG$ of a group $G$, denote by $\langle\mG\rangle_{grp}$ the sub\emph{group} generated by $\mG$.
\begin{enumerate}[noitemsep, label = (\roman*)]
    \setcounter{enumi}{\value{ProblemCounter}}
    \item \textit{(Group Membership)} given a finite set $\mG$ and an element $T$ in $G$, decide whether $T \in \langle\mG\rangle_{grp}$.
    \setcounter{ProblemCounter}{\value{enumi}}
\end{enumerate}
\cite{mikhailova1966occurrence} showed undecidability of Group Membership when $G$ is the group $\SL(4, \Z)$ of $4 \times 4$ integer matrices with determinant one.
One may note that undecidability of Group Membership subsumes that of Semigroup Membership by including the inverse of the elements in $\mG$.

There has been a steady growth in research intensity for Group and Semigroup Membership problems as they establish important connections between algebra and logic.
These problems now play an essential role in analysing system dynamics and program termination, and have numerous applications in automata theory, complexity theory, and interactive proof systems~\cite{beals1993vegas,blondel2005decidable,derksen2005quantum,hrushovski2018polynomial}.
It is worth noting that membership problems are in fact decidable for many classes of groups, such as abelian groups and low dimensional matrix groups~\cite{babai1996multiplicative,choffrut2005some}.
For example, in the matrix group $\SL(2, \Z)$, Semigroup Membership is decidable by a classic result of~\cite{choffrut2005some}, and Group Membership is decidable in polynomial time (PTIME) by a recent result of~\cite{lohrey2023subgroup}.
Our interest in computational group theory is two-fold.
From an application point of view, we are interested in developing practical algorithms for specific classes of groups.
From a theory point of view, we aim to close the gap between decidability and undecidability.

For most classes of groups, Group Membership is much more tractable than Semigroup Membership.
For example, Group Membership is decidable in the class of \emph{polycyclic groups}~\cite{kopytov1968solvability}; whereas Semigroup Membership is undecidable even in the subclass of \emph{nilpotent groups}~\cite{roman2022undecidability}.
This gap motivated the introduction of two intermediate problems in~\cite{choffrut2005some}:
\begin{enumerate}[noitemsep, label = (\roman*)]
    \setcounter{enumi}{\value{ProblemCounter}}
    \item \textit{(Identity Problem)} given a finite subset $\mG$ of $G$, decide whether $\sgmG$ contains the neutral element of $G$.
    \item \textit{(Group Problem)} given a finite subset $\mG$ of $G$, decide whether $\langle\mG\rangle = \langle\mG\rangle_{grp}$.
    \setcounter{ProblemCounter}{\value{enumi}}
\end{enumerate}
In other words, the Identity Problem asks whether there exists a non-empty sequence of elements from $\mG$ whose product is the neutral element; and the Group Problem asks whether the semigroup $\sgmG$ is a group.

The Group Problem is crucial in determining structural properties of a semigroup.
For example, given a decision procedure for the Group Problem, one can compute a generating set for the \emph{group of units} of a finitely generated semigroup $\sgmG$.
We also point out that there are significantly more available algorithms for groups than there are for semigroups.
Therefore performing preliminary checks using the Group Problem can help decide Semigroup Membership in many special cases.
Using the Group Problem, one can also decide the lesser known \emph{Inverse Problem}: given a finite set $\mG$ and an element $a \in \mG$, decide whether $a^{-1} \in \langle\mG\rangle$.
The solution for the Identity Problem is usually the most essential special case on the way to building an algorithm for Semigroup Membership.


Unfortunately, decidability of these two intermediate problems remains open for most classes of groups, even in cases where decidability of Group and Semigroup Membership already have definitive answers.
Notable examples include nilpotent groups, polycyclic groups and metabelian groups.

Beyond the membership problems and the intermediate problems, another classic problem is \emph{Semigroup Intersection}:
\begin{enumerate}[noitemsep, label = (\roman*)]
    \setcounter{enumi}{\value{ProblemCounter}}
    \item \textit{(Semigroup Intersection)} given two finite subsets $\mG, \mH$ of $G$, decide whether $\sgmG \cap \langle \mH \rangle = \emptyset$.
    \setcounter{ProblemCounter}{\value{enumi}}
\end{enumerate}
In the seminal paper where Markov demonstrated undecidabilty of Semigroup Membership, he also showed undecidability of Semigroup Intersection for integer matrix groups of dimension at least four~\cite{markov1947certain}.
Markov's idea is to encode the famous \emph{Post Correspondence Problem}, which can be reformulated as Semigroup Intersection in a direct product of two free monoids.

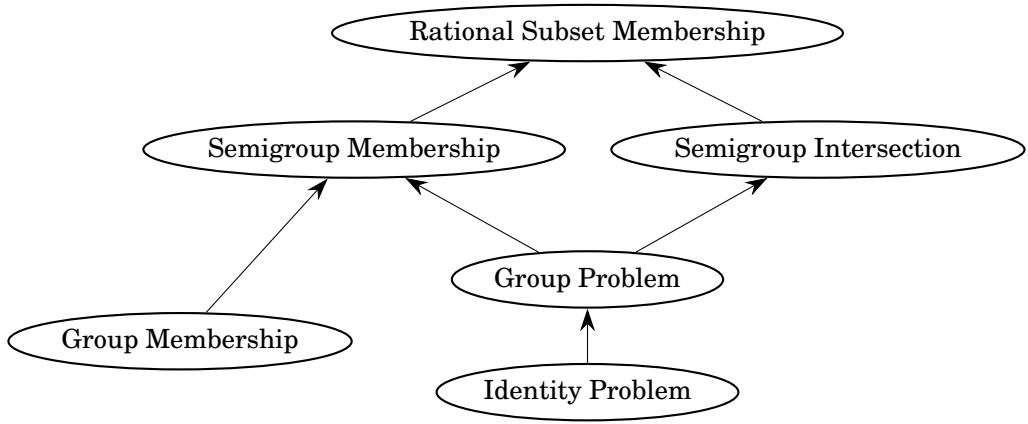
\begin{figure}[h]
    \centering
    \begin{tikzpicture}

        \node [draw, ellipse, thick, black, minimum width=1.5cm, minimum height=0.75cm, align=center, inner sep=1pt] (RSM) {Rational Subset Membership};

        \node [draw, ellipse, thick, black, minimum width=2cm, minimum height=0.75cm, inner sep=1pt, align=center, below left=1cm and -1.3cm of RSM] (SM) {Semigroup Membership};

        \draw [{Stealth[length=3mm, width=2mm]}-] (RSM) -- (SM);

        \node [draw, ellipse, thick, black, minimum width=2cm, minimum height=0.75cm, inner sep=1pt, align=center, below right=1cm and -1.3cm of RSM] (SI) {Semigroup Intersection};

        \draw [{Stealth[length=3mm, width=2mm]}-] (RSM) -- (SI);

        \node [draw, ellipse, thick, black, minimum width=2cm, minimum height=0.75cm, inner sep=1pt, align=center, below left=2cm and -1.3cm of SM] (GM) {Group Membership};

        \draw [{Stealth[length=3mm, width=2mm]}-] (SM) -- (GM);

        \node [draw, ellipse, thick, black, minimum width=2cm, minimum height=0.75cm, inner sep=1pt, align=center, below=2.5cm of RSM] (GP) {Group Problem};

        \draw [{Stealth[length=3mm, width=2mm]}-] (SM) -- (GP);
        \draw [{Stealth[length=3mm, width=2mm]}-] (SI) -- (GP);

        \node [draw, ellipse, thick, black, minimum width=2cm, minimum height=0.75cm, inner sep=1pt, align=center, below=4cm of RSM] (IP) {Identity Problem};

        \draw [{Stealth[length=3mm, width=2mm]}-] (GP) -- (IP);
        
    \end{tikzpicture}
    \medskip
    \caption{Reductions between different algorithmic problems.}
    \label{fig:reductions}
\end{figure}

We may note that some algorithmic problems are intrinsically more difficult than others.
For example, Semigroup Membership subsumes Group Membership by including the inverses of the generators.
In terms of decidability, both Semigroup Membership and Semigroup Intersection subsume the Group Problem, which itself subsumes the Identity Problem~\cite{bell2010undecidability}.
We also point out that all five problems are special cases of the more general \emph{Rational Subset Membership} problem, which asks whether an element $T$ is contained in a given \emph{rational subset} of $G$ (a subset defined by a rational expression over the generators of $G$).
Rational Subset Membership is beyond the scope of this article, and we refer interested readers to~\cite{lohrey2013rational} for a survey on this problem.
See also~\cite{lohrey2015rational,chistikov2016taming,DBLP:conf/icalp/CadilhacCZ20} for recent advances.
Reductions in terms of decidability between the different algorithmic problems is summarized in Figure~\ref{fig:reductions}.

\subsection*{Example: the case of $\Z^d$}
We end this introduction with an example to illustrate why the above algorithmic problems can be considered fundamental in computational group theory.

Let the ambient group $G$ be the abelian group $\Z^d$ for some $d \in \N$, where the group law ``$+$'' is coordinate-wise addition: $(a_1, \ldots, a_d) + (b_1, \ldots, b_d) = (a_1 + b_1, \ldots, a_d + b_d)$.
We show that the above algorithmic problems in the group $\Z^d$ correspond to several classic problems in computer science.

Consider Semigroup Membership in $\Z^d$. Let $\mG \coloneqq \{(a_{11}, \ldots, a_{1d}), \ldots, (a_{K1}, \ldots, a_{Kd})\}$ and $T \coloneqq (b_1, \ldots, b_d)$ be the input of Semigroup Membership. Recall that we want to decide whether $T$ is contained in the semigroup $\sgmG$.
Since $G = \Z^d$ is abelian, the semigroup $\sgmG$ can be explicitly described as follows:
\begin{equation}\label{eq:linearsg}
\sgmG = \left\{n_1 \cdot (a_{11}, \ldots, a_{1d}) + \cdots + n_K \cdot (a_{K1}, \ldots, a_{Kd}) \;\middle|\; (n_1, \ldots, n_K) \in \N^K \setminus \{0^K\}\right\}.
\end{equation}
Then, deciding whether $T \in \sgmG$ corresponds to deciding whether the system of linear equations
\begin{align*}
    & n_1 a_{11} + \cdots + n_K a_{K1} = b_1, \\
    & n_1 a_{12} + \cdots + n_K a_{K2} = b_2, \\
    & \quad \vdots \\
    & n_1 a_{1d} + \cdots + n_K a_{Kd} = b_d,
\end{align*}
admits a non-negative solution $(n_1, \ldots, n_K) \in \N^K \setminus \{0^K\}$.
This is equivalent to the fundamental problem of \emph{Integer Programming} and is therefore NP-complete.

Similarly, consider the problem of Group Membership in $\Z^d$, that is, whether $T$ is contained in the group $\gmG$ generated by the set $\mG$.
The group $\gmG$ can be explicitly described as follows:
\begin{equation}\label{eq:lineargr}
\gmG = \left\{n_1 \cdot (a_{11}, \ldots, a_{1d}) + \cdots + n_K \cdot (a_{K1}, \ldots, a_{Kd}) \;\middle|\; (n_1, \ldots, n_K) \in \Z^K\right\}.
\end{equation}
The difference of the \emph{group} $\gmG$ from the \emph{semigroup} $\sgmG$ is that we allow inverses of the generators to appear in the sum, hence the tuple $(n_1, \ldots, n_K)$ now takes value in $\Z^K$ instead of $\N^K \setminus \{0^K\}$.
Therefore, Group Membership is equivalent to deciding whether the system of linear equations
\begin{align*}
    & n_1 a_{11} + \cdots + n_K a_{K1} = b_1, \\
    & n_1 a_{12} + \cdots + n_K a_{K2} = b_2, \\
    & \quad \vdots \\
    & n_1 a_{1d} + \cdots + n_K a_{Kd} = b_d,
\end{align*}
admits an integer solution $(n_1, \ldots, n_K) \in \Z^K$.
This becomes the classic problem of solving linear equations over $\Z$, and is decidable in polynomial (less then cubic) time.

Consider now the Identity Problem in $\Z^d$. Specializing Semigroup Membership by taking $T$ to be the neutral element $(0, \ldots, 0)$, the Identity Problem is equivalent to deciding whether the system of \emph{homogeneous} linear equations
\begin{align*}
    & n_1 a_{11} + \cdots + n_K a_{K1} = 0, \\
    & n_1 a_{12} + \cdots + n_K a_{K2} = 0, \\
    & \quad \vdots \\
    & n_1 a_{1d} + \cdots + n_K a_{Kd} = 0,
\end{align*}
admits a non-trivial solution $(n_1, \ldots, n_K) \in \N^K \setminus \{0^K\}$.
By scaling the solution, this is equivalent to deciding whether the system admits a non-trivial solution over the non-negative \emph{rationals} $(n_1, \ldots, n_K) \in \Q_{\geq 0}^K \setminus \{0^K\}$.
Since the system only admits integer coefficients, it admits a non-trivial solution over the non-negative rationals if and only if it admits a non-trivial solution over the non-negative \emph{reals}.
Hence, the Identity Problem becomes the famous \emph{Linear Programming} problem, which is known to be decidable in polynomial time.

The Group Problem is perhaps less intuitive in terms of systems of linear equations. However it becomes clearer from a geometric point of view.
Indeed, the semigroup $\sgmG$ can be understood as the cone generated by the elements $\mG$ in the space $\Z^d$, whereas the group $\gmG$ can be understood as the linear space generated by the elements $\mG$.
Therefore the Group Problem can be reformulated as deciding whether a finitely generated cone is actually a linear space.

For Semigroup Intersection,
write $\mG \coloneqq \{(a_{11}, \ldots, a_{1d}), \ldots, (a_{K1}, \ldots, a_{Kd})\}$ and $\mH \coloneqq \{(c_{11}, \ldots, c_{1d}), \ldots, (c_{M1}, \ldots, c_{Md})\}$.
Then deciding whether $\sgmG \cap \langle \mH \rangle = \emptyset$ is equivalent to deciding whether the system of homogeneous linear equations
\begin{align*}
    & n_1 a_{11} + \cdots + n_K a_{K1} = \ell_1 c_{11} + \cdots + \ell_M c_{M1}, \\
    & n_1 a_{12} + \cdots + n_K a_{K2} = \ell_1 c_{12} + \cdots + \ell_M c_{M2}, \\
    & \quad \vdots \\
    & n_1 a_{1d} + \cdots + n_K a_{Kd} = \ell_1 c_{1d} + \cdots + \ell_M c_{Md},
\end{align*}
admits a pair of non-trivial solutions $(n_1, \ldots, n_K) \in \N^{K} \setminus \{0^{K}\}, (\ell_1, \ldots, \ell_M) \in \N^{M} \setminus \{0^{M}\}$.
This is again solvable by Linear Programming by scaling the solutions to $\Q_{\geq 0}$.
In particular, since we are working in the abelian group $\Z^d$, commutativity allows us to freely permute elements in the sum, and move terms between the two sides of an equation.
Hence Semigroup Intersection becomes similar to the Identity Problem.
However, when we work in groups with more complex structures (namely non-abelian groups), this equivalence will fail, and Semigroup Intersection can become much more difficult than the Identity Problem.

Lastly, we mention that Rational Subset Membership in $\Z^d$ corresponds to the notion of computing \emph{semilinear sets}.
See~\cite{chistikov2016taming} for a proof of decidability and its complexity analysis.

\section{Notes for the reader}
It is clear from the introduction that the complexity and decidability of semigroup algorithmic problems greatly depend on the ambient group $G$.
As shown in the last example, the case where $G = \Z^d$ is relatively simple.
On the contrary, when the group $G$ is an arbitrary (non-commutative) group, the algorithmic problems we consider become much more complex and even undecidable.
To achieve decidability, one must assume additional structure on the group $G$.
The aim of this article is to give an overview of recent advances in the decidability and complexity of the aforementioned algorithmic problems in different classes of groups $G$, as well as point out various open problems.
Section~\ref{sec:mat} is about low-dimensional matrix groups as well as their subgroups and extensions, whereas in Section~\ref{sec:abs} we consider groups with additional structures such as nilpotency and being metabelian.
In Section~\ref{sec:other} we mention some other related algorithmic problems in groups and semigroups.


For the most part of this article, the group $G$ is explicitly given as a matrix group (such as $\SL(n, \Z)$). In this case, the elements of $G$ are represented as matrices with \emph{binary encoded entries}.
Nevertheless, many of the results stated in this article can be extended to the case where $G$ is given as an abstract group, which is usually embeddable in some well-studied matrix group.

\section{Low-dimensional matrix groups}\label{sec:mat}
This section contains results for low-dimensional matrix groups.
In particular, we are interested in the \emph{special linear groups} $\SL(n, \Z)$ as well as their subgroups and extensions.

\begin{definition}[Special linear groups]\label{def:SL}
Let $n \in \N$. The \emph{special linear group} $\SL(n, \Z)$ is the group of $n \times n$ integer matrices with determinant one:
\[
\SL(n, \Z) \coloneqq \{A \in \Z^{n \times n} \mid \det(A) = 1\}.
\]
\end{definition}


Semigroups are closely related to the notion of \emph{words} over an alphabet:
\begin{definition}[Words over an alphabet]
    By an \emph{alphabet}, we mean a finite set $\Sigma$.
    Elements of an alphabet are called \emph{letters}.
    A \emph{word} over an alphabet $\Sigma$ is a finite string of letters, possibly empty.
    In particular, the empty string is called the \emph{empty word}, usually denoted by $\epsilon$.
    Given any alphabet $\Sigma$, we denote by $\Sigma^*$ the set of words over $\Sigma$:
    \[
    \Sigma^* \coloneqq \left\{a_{1} a_{2} \cdots a_{m} \mid m \geq 0, a_{1}, \ldots a_{m} \in \Sigma \right\}.
    \]
    For two words $v, w \in \Sigma^*$, we denote by $v \cdot w$ or simply $v w$ the concatenation of $v$ and $w$, it is again a word over $\Sigma$.
    The operation of concatenation gives the sets $\Sigma^*$ and $\Sigma^* \setminus \{\epsilon\}$ the structure of semigroups.
\end{definition}

For example, Semigroup Membership can now be reformulated as asking whether there exists a word $w \in \mG^* \setminus \{\epsilon\}$, whose product in $G$ is equal to $T$; and Semigroup Intersection can be reformulated as asking whether there exists two words $w \in \mG^* \setminus \{\epsilon\}, v \in \mH^* \setminus \{\epsilon\}$, such that the products of $w$ and $v$ in $G$ are equal.

\subsection{Dimension two}

The group $\SL(2, \Z)$ has been a classic object of study. It is closely related to various areas in mathematics such as hyperbolic geometry, modular forms, and dynamical systems~\cite{bruinier2008elliptic,newman1962structure,polterovich2004stable}. 

The key to solving algorithmic problems in $\SL(2, \Z)$ is the fact that $\SL(2, \Z)$ is \emph{virtually free}, meaning it contains a \emph{free} subgroup of finite index.

\begin{definition}[Free group]\label{def:free}
    Given an alphabet $\Sigma$, define the corresponding group alphabet $\Sigma^{\pm} \coloneqq \Sigma \cup \{a^{-1} \mid a \in \Sigma\}$, where $a^{-1}$ is a new letter for each $a \in \Sigma$.
    There is a natural involution $(\cdot)^{-1}$ over the set of words $\left(\Sigma^{\pm}\right)^*$ defined by $(a^{-1})^{-1} = a$ and $(a_1 a_2 \cdots a_m)^{-1} = a_m^{-1} \cdots a_2^{-1} a_1^{-1}$. 
    A word over the alphabet $\Sigma^{\pm}$ is called \emph{reduced} if it does not contain consecutive letters $a a^{-1}$ or $a^{-1} a$, $a \in \Sigma$.
    For a word $w$, define $\operatorname{red}(w)$ to be the reduced word obtained by iteratively replacing consecutive letters $a a^{-1}$ and $a^{-1} a$ with the empty string.
    The \emph{free group} $F(\Sigma)$ over $\Sigma$ is then defined as the set of reduced words in $\left(\Sigma^{\pm}\right)^*$, where multiplication is given by $v \cdot w = \operatorname{red}(vw)$, inversion is given by the involution $(\cdot)^{-1}$, and the neutral element is the empty word $\epsilon$.
\end{definition}

If we take $\Sigma = \{a, b\}$, then the group $F(\Sigma)$ is the free group over two generators, and is denoted by $F_2$.
For example, in $F(\{a, b\})$ we have $(aaba^{-1})^{-1} = a b^{-1} a^{-1} a^{-1}$, and $aaba^{-1} \cdot a b = \operatorname{red}(aaba^{-1} a b) = aabb$.

In particular, the subgroup of $\SL(2, \Z)$ generated by 
$
A \coloneqq
\begin{pmatrix}
    1 & 2 \\
    0 & 1 \\
\end{pmatrix}
$
and
$
B \coloneqq
\begin{pmatrix}
    1 & 0 \\
    2 & 1 \\
\end{pmatrix}
$
is a free group~\cite[Example~7.63]{dructu2018geometric}. It is of index 12, meaning the quotient $\SL(2, \Z) / \langle A, B \rangle_{grp}$ is a set with 12 elements~\cite{43740}.

Algorithmic problems in free groups have been extensively studied since the 1950s~\cite{howson1954intersection}.
A classic tool for solving algorithmic problem in free groups is the automata-inspired \emph{Stallings foldings}.
The idea of using automata-based approaches to deal with free groups can be traced back to~\cite{benois1969parties} who showed decidability of Rational Subset Membership in finitely generated free groups.
We recommend~\cite{KAPOVICH2002608} for an introduction on Stallings foldings as well as their applications in algorithmic problems in free groups.

While Stallings foldings are initially proposed to deal with free groups, they are easily generalizable to virtually free groups, notably $\SL(2, \Z)$.
This becomes the foundation of several results concerning decidability and complexity of algorithmic problems in $\SL(2, \Z)$:

\begin{theorem}
    In the group $\SL(2, \Z)$:
    \begin{enumerate}[nosep, label = (\roman*)]
        \item \cite{choffrut2005some} Rational Subset Membership is decidable.
        \item \cite{bell2017identity,bell2023private} Semigroup Membership, the Identity Problem and the Group Problem are decidable and NP-complete.
        \item \cite{lohrey2023subgroup} Group Membership is decidable in PTIME.
    \end{enumerate}
\end{theorem}

\subsection*{Example: Semigroup Membership in the free group $F_2$}

We give an example of the decision procedure for Semigroup Membership in the free group $F_2$ to show some of the techniques involved.
This is very similar to the decision procedure for $\SL(2, \Z)$ (which contains $F_2$ as a finite index subgroup), but is easier to describe.

Working in the group $F_2 = F(\{a, b\})$, suppose we want to decide whether $T \coloneqq a$ is in the semigroup generated by the set $\mG \coloneqq \{aab, b^{-1} a^{-1}\}$.
In other words, we want to decide whether there is an expression of the form $w = w_1 w_2 \cdots w_p, p \geq 1, w_1, \ldots, w_p \in \{aab, b^{-1} a^{-1}\}$, such that $\operatorname{red}(w) = a$ (recall the definition of $\operatorname{red}(\cdot)$ from Definition~\ref{def:free}).
Since $a \neq \epsilon$, we can without loss of generality include the case $p = 0$.
Therefore, deciding whether $a \in \langle aab, b^{-1} a^{-1} \rangle$ is equivalent to deciding whether there exists an expression of the form $v = a^{-1} w_1 w_2 \cdots w_p,\; p \geq 0, w_1, \ldots, w_p \in \{aab, b^{-1} a^{-1}\}$, such that $\operatorname{red}(v) = \epsilon$.

We can construct an automaton $\mA$ that recognizes the set of words
\[
\mL \coloneqq \big\{a^{-1} w_1 w_2 \cdots w_p \;\big|\; p \geq 0, w_1, \ldots, w_p \in \{aab, b^{-1} a^{-1}\} \big\}.
\]
See Figure~\ref{fig:automaton} for an illustration of the automaton $\mA$.
In particular, every path starting from the initial state $q_I$ and ending at the accepting state $q_F$ represents a word in the language $\mL$; and every word in $\mL$ can be represented as a path from $q_I$ to $q_F$.
For example, the word $a^{-1} \cdot aab \cdot b^{-1} a^{-1} \in \mL$ corresponds to the path $q_I \rightarrow q_F \rightarrow q_{11} \rightarrow q_{12} \rightarrow q_{F} \rightarrow q_{21} \rightarrow q_{F}$.

\tikzset{every state/.style={inner sep=2pt, minimum size=0pt}}
\begin{figure}[h]
    \centering
    \begin{minipage}{.45\textwidth}
        \begin{tikzpicture}[shorten >=1pt,node distance=2cm,on grid,auto, inner sep=1pt] 
            \node[state,initial] (q_I)   {$q_I$}; 
            \node[state,accepting] (q_F) [right= 3cm of q_I] {$q_F$};
            \node[state] (q_11) [above left= 2cm and 1cm of q_F] {$q_{11}$};
            \node[state] (q_12) [above right= 2cm and 1cm of q_F] {$q_{12}$};
            \node[state] (q_21) [below= 2cm of q_F] {$q_{21}$};   
            \path[->] 
            (q_I) edge [bend right] node {$a^{-1}$} (q_F)
            (q_F) edge [bend left] node {$a$} (q_11)
            (q_11) edge [bend left] node {$a$} (q_12)
            (q_12) edge [bend left] node {$b$} (q_F)
            (q_F) edge [bend left] node {$b^{-1}$} (q_21)
            (q_21) edge [bend left] node {$a^{-1}$} (q_F)
        ;
        \end{tikzpicture}
        \caption{An automaton $\mA$ that recognizes the language $\mL$.}
        \label{fig:automaton}
    \end{minipage}%
    \hfill
    \begin{minipage}{.45\textwidth}
        \begin{tikzpicture}[shorten >=1pt,node distance=2cm,on grid,auto, inner sep=1pt] 
            \node[state,initial] (q_I)   {$q_I$}; 
            \node[state,accepting] (q_F) [right= 3cm of q_I] {$q_F$};
            \node[state] (q_11) [above left= 2cm and 1cm of q_F] {$q_{11}$};
            \node[state] (q_12) [above right= 2cm and 1cm of q_F] {$q_{12}$};
            \node[state] (q_21) [below= 2cm of q_F] {$q_{21}$};   
            \path[->] 
            (q_I) edge [bend right] node {$a^{-1}$} (q_F)
            (q_F) edge [bend left] node {$a$} (q_11)
            (q_11) edge [bend left] node {$a$} (q_12)
            (q_12) edge [bend left] node {$b$} (q_F)
            (q_F) edge [bend left] node {$b^{-1}$} (q_21)
            (q_21) edge [bend left] node {$a^{-1}$} (q_F)
            (q_21) edge [dashed, bend left] node[right] {$\epsilon$} (q_11)
            (q_12) edge [dashed, bend left] node {$\epsilon$} (q_21)
            (q_11) edge [dashed, bend left] node {$\epsilon$} (q_F)
            (q_I) edge [dashed, bend left] node {$\epsilon$} (q_11)
            (q_I) edge [dashed, bend left] node {$\epsilon$} (q_F)
        ;
        \end{tikzpicture}
        \caption{The saturated automaton $\mA_0$.}
        \label{fig:automatonsat}
    \end{minipage}%
\end{figure}

However, it is not clear from the automaton $\mA$ whether $\mL$ contains a word $v$ such that $\operatorname{red}(v) = \epsilon$.
The idea is to now ``saturate'' the automaton $\mA$ to include words obtained from reducing words in $\mL$.
To be precise, we want to construct an automaton $\mA_0$ that recognizes a language $\mL_0$, such that the reduced words in $\mL_0$ are exactly the set $\{\operatorname{red}(v) \mid v \in \mL\}$.

A construction of the automaton $\mA_0$ is illustrated in Figure~\ref{fig:automatonsat}.
Starting with $\mA$, we add the edge $q_{12} \xrightarrow{\epsilon} q_{21}$ because any path containing the subpath $q_{12} \xrightarrow{b} q_{F} \xrightarrow{b^{-1}} q_{21}$ results in $bb^{-1}$ which can be reduced to $\epsilon$.
Similarly, we add the edge $q_{11} \xrightarrow{\epsilon} q_{F}$ thanks to the path $q_{11} \xrightarrow{a} q_{12} \xrightarrow{\epsilon} q_{21} \xrightarrow{a^{-1}} q_{F}$ resulting in $aa^{-1}$; we add the edge $q_{21} \xrightarrow{\epsilon} q_{11}$ thanks to the path $q_{21} \xrightarrow{a^{-1}} q_{F} \xrightarrow{a} q_{11}$; and we add the edge $q_{I} \xrightarrow{\epsilon} q_{11}$ thanks to the path $q_{I} \xrightarrow{a^{-1}} q_{F} \xrightarrow{a} q_{11}$.
Finally, we add the edge $q_{I} \xrightarrow{\epsilon} q_{F}$ thanks to the path $q_{I} \xrightarrow{\epsilon} q_{11} \xrightarrow{\epsilon} q_{F}$.
As a conclusion, we can pass from $q_I$ to $q_F$ using a path which reduces to $\epsilon$.
Therefore $T \coloneqq a$ is indeed contained in the semigroup generated by the set $\mG \coloneqq \{aab, b^{-1} a^{-1}\}$.

Note that this process of adding $\epsilon$-transitions always finishes in finitely many steps because there are only a finite number of states in $\mA$.
In general, if after the process there is an $\epsilon$-transition from $q_I$ to $q_F$ then we conclude that $T \in \sgmG$; otherwise we conclude $T \not\in \sgmG$.
Note that the termination of this process relies on the fact that $\operatorname{red}(\cdot)$ is \emph{deterministic}, meaning for every $v$ the image $\operatorname{red}(v)$ is unique.
This feature grants free groups their relatively simple structure as well as decidability of their various algorithmic problems.

The group $\SL(2, \Z)$ is not far from being a free group, and a similar approach to the above automata construction can also be applied to $\SL(2, \Z)$.
We refer readers to~\cite{potapov2017decidability} for a detailed account of the method for $\SL(2, \Z)$.

\subsection*{Semigroup Membership in extensions of $\SL(2, \Z)$}

Note that one can naturally generalize the definition of Semigroup Membership to the case where the ambient group $G$ is a \emph{semigroup} instead of a group.
A natural follow-up question to $\SL(2, \Z)$ is whether Semigroup Membership remains decidable when $G$ is the semigroup $\Z^{2 \times 2}$ of $2 \times 2$ integer matrices.
Compared to $\SL(2, \Z)$, the semigroup $\Z^{2 \times 2}$ contains matrices with determinant $0$ and $\pm d, d \geq 1$, which add an extra layer of difficulty.
Nevertheless, several important partial results have been obtained via extensions of the solution for $\SL(2, \Z)$:

\begin{theorem}
    Semigroup Membership is decidable in:
    \begin{enumerate}[nosep, label = (\roman*)]
        \item \cite{potapov2017decidability} the set of $2 \times 2$ integer matrices with non-zero determinants.
        \item \cite{DBLP:conf/mfcs/PotapovS17} the set of $2 \times 2$ integer matrices with determinants $0, -1$ and $1$.
    \end{enumerate}
\end{theorem}

Decidability of Semigroup Membership remains an open problem for the set of \emph{all} $2 \times 2$ integer matrices.

Another interesting extension of $\SL(2, \Z)$ is the group $\GL(2, \Q)$ of $2 \times 2$ invertible matrices with \emph{rational} entries.
All algorithmic problems listed in Figure~\ref{fig:reductions} remain open for the group $\GL(2, \Q)$, with very limited partial results currently known~\cite{diekert2020decidability,DBLP:conf/icalp/CadilhacCZ20}.

\subsection{Dimension three}\label{subsec:dimthree}
One of the most outstanding open problems in computational group theory is the decidability of Group Membership in $\SL(3, \Z)$.
In fact, the decidability status of all algorithmic problems listed in Figure~\ref{fig:reductions} remains open for $\SL(3, \Z)$.
This is notably due to a lack of understanding for the structure of subgroups in $\SL(3, \Z)$.
Currently, the closest undecidability result is Semigroup Membership in the set $\Z^{3 \times 3}$ of $3 \times 3$ integer matrices, by~\cite{paterson1970unsolvability}:

\begin{theorem}[Mortality Problem~\cite{paterson1970unsolvability}] The following problem is undecidable: given a finite set $\mG$ of $3 \times 3$ integer matrices, decide whether the zero matrix $0^{3 \times 3}$ is in the semigroup $\sgmG$.
\end{theorem}

Although the decidability of algorithmic problems in $\SL(3, \Z)$ seems out of reach for the moment, positive results have been obtained for several important subgroups of $\SL(3, \Z)$.

\subsection*{The Heisenberg group $\HH_3(\Z)$}

We consider the \emph{integer Heisenberg group} $\HH_3(\Z)$, which is the group of $3 \times 3$ upper-triangular integer matrices with ones on the diagonal.
\[
\HH_3(\Z) \coloneqq 
\left\{
\begin{pmatrix}
1 & * & * \\
0 & 1 & * \\
0 & 0 & 1 \\
\end{pmatrix}
, \text{ where $*$ are entries in $\Z$} \right\},
\]

The Heisenberg groups play an important role in many branches of mathematics, physics and computer science.
They first arose in the description of one-dimensional quantum mechanical systems~\cite{Neumann1931,weyl1950theory}, and have now become an important mathematical object connecting areas like representation theory, Fourier analysis and quantum algorithms~\cite{howe1980role,kirillov2004lectures,krovi2008efficient}.
From a computational point of view, the Heisenberg group received much attention in the past ten years because it is one of the simplest non-commutative matrix groups.

\begin{theorem}\label{thm:heis}
    In the Heisenberg group $\HH_3(\Z)$:
    \begin{enumerate}[nosep, label = (\roman*)]
        \item \cite{colcombet2019reachability} Semigroup Membership is decidable.
        \item \cite{ko2017identity} The Identity Problem is decidable in PTIME.
        \item \cite{dong2022identity} The Group Problem is decidable in PTIME.
        \item \cite{DBLP:conf/stacs/000123} Semigroup Intersection is decidable in PTIME.
    \end{enumerate}
\end{theorem}
It follows from Theorem~\ref{thm:heis}(i) that Group Membership is also decidable in $\HH_3(\Z)$.
However we are not aware of any complexity analysis on this problem.
The main idea for proving Theorem~\ref{thm:heis}(i) is to use the \emph{Baker-Campbell-Hausdorff formula}~\cite{baker1905alternants,campbell1897law,hausdorff1906symbolische} from Lie algebra, as well as to incorporate Semigroup Membership in a Parikh automaton.
The main ideas behind Theorem~\ref{thm:heis}(ii)-(iv) are reductions to word combinatorics problems.
Note that the decidability of Rational Subset Membership in $\HH_3(\Z)$ remains an intricate open problem.

\subsection*{The special affine group $\SA$}

Compared to $\HH_3(\Z)$, a larger subgroup of $\SL(3, \Z)$ is the \emph{special affine group} $\SA$.
The group $\SA$ consists of affine transformations of $\Z^2$ that preserve orientation, and can be considered as an intermediate group between $\SL(2, \Z)$ and $\SL(3, \Z)$.
Written as matrices, elements of $\SA$ are $3 \times 3$ integer matrices of the following form.
\[
\SA \coloneqq \left\{ 
M = 
\begin{pmatrix}
* & * & * \\
* & * & * \\
0 & 0 & 1 \\
\end{pmatrix}
\;\middle|\;
\det(M) = 1 \text{, where $*$ are entries in $\Z$}
\right\}.
\]
Like $\SL(2, \Z)$, the special affine group is important in the context of many fundamental problems, such as Lie groups~\cite{wolf1963affine}, polyhedral geometry~\cite{mundici2014invariant}, dynamical systems~\cite{cabrer2017classifying}, and computer vision~\cite{giefer2020extended}.
Apart from the intrinsic interest to study $\SA$, we also point out that the Special Affine group has tight connections to various reachability problems in automata theory.
Some of the central questions in automated verification include reachability problems in \emph{Affine Vector Addition Systems} and \emph{Affine Vector Addition Systems with states (Affine VASS)} over the integers~\cite{raskin2021affine}.
The study of these reachability problems in dimension two necessitates the understanding of subsemigroups of $\SA$.

Every element in $\SA$ can be written as a pair $(A, \ba)$, where $A \in \SL(2, \Z), \ba \in \Z^2$.
This pair represents the element 
$\begin{pmatrix}
        A & \ba \\
        0 & 1 \\
\end{pmatrix}$.
Multiplication in $\SA$ is then given by
\[
(A, \ba) \cdot (B, \boldsymbol{b}) = (AB, A \boldsymbol{b} + \ba).
\]
Naturally, $\SA$ has a subgroup $\{(A, \bzer) \mid A \in \SL(2, \Z)\}$ isomorphic to $\SL(2, \Z)$.
In the language of group theory, this means that $\SA$ is a \emph{semidirect product} of the groups $\SL(2, \Z)$ and $\Z^2$.
This structure is crucial to the resolution of several algorithmic problems in $\SA$.

\begin{theorem}\label{thm:SA}
    In the group $\SA$:
    \begin{enumerate}[nosep, label = (\roman*)]
        \item \cite{kapovich2005foldings,delgado2017extensions} Group Membership is decidable.
        \item \cite{dong2023identity} The Identity Problem and the Group Problem are decidable and NP-complete.
    \end{enumerate}
\end{theorem}

The decidability of Semigroup Membership and Semigroup Intersection in $\SA$ remains open.
See~\cite{dong2023identity} for a discussion of current obstacles to solving these problems.
Theorem~\ref{thm:SA}(i) comes from the classic idea that $\SA$ can be realized as a \emph{graph of groups}.
Graph of groups is a central object in the area of geometric group theory, and Group Membership has been solved for a large class of graph of groups using generalizations of Stallings foldings~\cite{kapovich2005foldings}.
However, such methods generally fail for semigroup problems, and the proof of Theorem~\ref{thm:SA}(ii) employs an algebraic approach instead of an automata-based approach.

\subsection{Dimension four}\label{subsec:dimfour}
While the algorithmic problems of Figure~\ref{fig:reductions} are decidable in $\SL(2, \Z)$ and open in $\SL(3, \Z)$, they all become undecidable for $\SL(4, \Z)$.
Note that $\SL(4, \Z)$ naturally contains as a subgroup the direct product $\SL(2, \Z) \times \SL(2, \Z)$, and that $\SL(2, \Z)$ contains the free subgroup $F_2$. 
Therefore $\SL(4, \Z)$ contains the subgroup $F_2 \times F_2$, which constitutes the key to proving all undecidability results in $\SL(4, \Z)$.
If a problem is undecidable in $F_2 \times F_2$, then it must also be undecidable in $\SL(4, \Z)$.

\begin{theorem}\label{thm:SL4}
    In the groups $F_2 \times F_2$ and $\SL(4, \Z)$:
    \begin{enumerate}[nosep, label = (\roman*)]
        \item \cite{mikhailova1966occurrence} Group Membership is undecidable.
        \item \cite{bell2010undecidability} The Identity Problem is undecidable.
    \end{enumerate}
\end{theorem}
As a result of Theorem~\ref{thm:SL4}, all algorithmic problems listed in Figure~\ref{fig:reductions} are undecidable in $F_2 \times F_2$ and in $\SL(4, \Z)$.
Mikhailova's idea of proving Theorem~\ref{thm:SL4}(i) is an embedding of the \emph{Word Problem}, which is one of the three fundamental problems of computational group theory.

\begin{theorem}[{The Word Problem~\cite{novikov1955algorithmic}}]\label{thm:wordprob}
    The following problem is undecidable:
    given an alphabet $\Sigma$ as well as elements $v_1, \ldots, v_n, w$ in the free group $F(\Sigma)$, decide whether $w$ is in the normal subgroup $\langle\langle v_1, \ldots, v_n \rangle\rangle$ generated by $v_1, \ldots, v_n$.
    In other words, it is undecidable whether a word $w \in \left(\Sigma^{\pm}\right)^*$ is equal to the neutral element in the quotient $F(\Sigma) /\langle\langle v_1, \ldots, v_n\rangle\rangle$.
\end{theorem}

The subgroup $\langle\langle v_1, \ldots, v_n\rangle\rangle$ is defined as the smallest \emph{normal} subgroup containing the elements $v_1, \ldots, v_n$. It is in general different from $\langle v_1, \ldots, v_n \rangle_{grp}$. This constitutes the difference between the Word Problem and Group Membership.
For this reason, Group Membership is sometimes referred to in literature as the \emph{Generalized Word Problem}.
Mikhailova showed that one can embed an instance of the Word Problem into the Group Membership problem in $F_2 \times F_2$, thus proving its undecidability.

Bell and Potapov's idea of proving Theorem~\ref{thm:SL4}(ii) is to embed a variation of the famous \emph{Post Correspondence Problem}.
\begin{theorem}[{Post Correspondence Problem~\cite{bams/1183507843}}]\label{thm:PCP}
    The following problem is undecidable:
    given a set of word pairs $S = \{(v_1, w_1), \ldots, (v_K, w_K)\}$ in $\{a, b\}^* \times \{a, b\}^*$, decide whether there exist $p \geq 1$ and a sequence of indices $i_1, \ldots, i_p \in \{1, \ldots, K\}$ such that $v_{i_1} v_{i_2} \cdots v_{i_p} = w_{i_1} w_{i_2} \cdots w_{i_p}$. 
\end{theorem}
The Post Correspondence Problem is one of the first known undecidability results in algorithmic theory, and it remains a staple tool for proving undecidability of various combinatorial problems.
Bell and Potapov showed that one can embed an instance of the Post Correspondence Problem into the Identity Problem in $F_2 \times F_2$, thus proving its undecidability.

\begin{remark}\label{rmk:embedPCP}
    The Post Correspondence Problem can be considered as a special case of
    Semigroup Intersection in the semigroup $\{a, b\}^* \times \{a, b\}^*$.
    Indeed, the Post Correspondence Problem can be reformulated as deciding whether the semigroup $\langle (v_1, w_1), \ldots, (v_K, w_K) \rangle \subseteq \{a, b\}^* \times \{a, b\}^*$ intersects the semigroup of diagonals $\langle (a, a), (b, b) \rangle$. 
    Therefore, if a group $G$ contains $\{a, b\}^* \times \{a, b\}^*$ as a subsemigroup, then $G$ has undecidable Semigroup Intersection.
    Since the group $F_2 = F(\{a, b\})$ naturally contains the subsemigroup $\{a, b\}^*$, we immediately obtain undecidability of Semigroup Intersection in $F_2 \times F_2$.
    Bell and Potapov's main contribution for Theorem~\ref{thm:SL4}(ii) is the strengthening of the undecidability result from Semigroup Intersection to the Identity Problem.

    It is worth pointing out that the same method cannot be applied to the group $\SL(3, \Z)$.
    Indeed, it has been shown in~\cite[Theorem~11]{ko2017identity} that $\SL(3, \Z)$ does not contain any subsemigroup isomorphic to $\{a, b\}^* \times \{a, b\}^*$.
    This added an extra layer of mystery to the decidability status of problems in $\SL(3, \Z)$.
\end{remark}

\section{Groups with additional structure}\label{sec:abs}
In Section~\ref{sec:mat} we showed that most algorithmic problems that are undecidable for general matrix groups become decidable when restricted to low dimensions.
In this section, instead of the restriction on dimension, we consider restrictions on the \emph{structure} of the matrix group.
An iconic result in this direction comes from Babai, Beals, Cai, Ivanyos and Luks~\cite{babai1996multiplicative}, who showed the NP-completeness of Semigroup Membership and PTIME decidability of Group Membership in \emph{commutative} matrix groups.
From the introduction (Section~\ref{sec:intro}), we also know that the problems in Figure~\ref{fig:reductions} all have classic interpretations in the commutative group $\Z^d$.
In this section we consider various relaxations of the commutativity requirement.

Recall from Subsection~\ref{subsec:dimfour} that all problems in Figure~\ref{fig:reductions} become undecidable when the group $G$ contains the direct product $F_2 \times F_2$ of two free groups.
Most natural classes of groups are stable under direct product (for example, the direct product of two commutative groups is still commutative).
This motivates us to consider classes of groups that do not contain $F_2$ as subgroup, as these are the only natural candidates for positive decidability results.

By the following celebrated result of Tits, a matrix group that does not contain the subgroup $F_2$ must be \emph{virtually solvable}, meaning it contains a \emph{solvable} subgroup of finite index.

\begin{theorem}[{Tits alternative~\cite{TITS1972250}}]\label{thm:Tits}
    Fix any $n \in \N$ and let $\K$ be any field.
    Recall that $\GL(n, \K)$ denotes the group of $n \times n$ invertible matrices with entries in $\K$.
    For every finitely generated subgroup $H$ of $\GL(n, \K)$, exactly one of the following is true:
    \begin{enumerate}[nosep, label = (\roman*)]
        \item $H$ contains $F_2$ as a subgroup.
        \item $H$ contains a solvable subgroup of finite index.
    \end{enumerate}
\end{theorem}

We therefore concentrate our study on the class of solvable groups.
Its exact definition is given as follows.

\begin{definition}[Solvable groups]\label{def:solvable}
Given a group $G$ and its subgroup $H$,
define the commutator $[G, H]$ to be the group generated by the elements $\{ghg^{-1}h^{-1} \mid g \in G, h \in H\}$.
The \emph{derived series} of a group $G$ is the inductively defined descending sequence of subgroups
\[
G = G^{(0)} \geq G^{(1)} \geq G^{(2)} \geq \cdots,
\]
in which $G^{(k)} = [G^{(k-1)}, G^{(k-1)}]$.
A group $G$ is called \emph{solvable} if its derived series terminates with $G^{(d)}$ being the trivial group for some finite $d$.
In this case, the smallest such $d$ is called the \emph{derived length} of $G$.
For example, an abelian group is solvable of derived length one.
\end{definition}

We refer interested readers to~\cite[Chapter~13]{dructu2018geometric} for a comprehensive introduction of solvable groups.
It is well known that all subgroups, quotients and direct products of solvable groups are solvable~\cite[Proposition~13.91]{dructu2018geometric}.

\begin{example}
    Denote by $\T(n, \Q)$ the group of $n \times n$ invertible upper triangular matrices with rational entries:
    \[
    \T(n, \Q) \coloneqq 
    \left\{
    \begin{pmatrix}
    a_1 & * & \cdots & * & * \\
    0 & a_2 & \cdots & * & * \\
    \vdots & \vdots & \ddots & \vdots & \vdots \\
    0 & 0 & \cdots & a_{n-1} & * \\
    0 & 0 & \cdots & 0 & a_n \\
    \end{pmatrix}
    , \text{ where $a_i$ and $*$ are entries in $\Q$ }, a_1 a_2 \cdots a_n \neq 0 \right\}.
    \]
    Then the group $\T(n, \Q)$ is solvable.
\end{example}

Unfortunately, without additional constraints, algorithmic problems in solvable groups remain highly intractable.
For example, \cite{kharlampovich1981finitely} famously constructed a solvable group of derived length three with undecidable Word Problem\footnote{To be precise, this means that the undecidability result in Theorem~\ref{thm:wordprob} still holds when $\Sigma, v_1, \ldots, v_n$ are fixed so that the quotient $F(\Sigma) /\langle\langle v_1, \ldots, v_n\rangle\rangle$ is a solvable group of derived length three.}.
Other problems such as Semigroup Membership and Semigroup Intersection are also undecidable in general solvable groups (see Subsections~\ref{subsec:nilp} and \ref{subsec:meta}).
Therefore, in order to obtain decidability results, we want to consider solvable groups with additional constraints.
In what follows, we consider two important special classes of solvable groups: \emph{nilpotent groups} and \emph{metabelian groups}.

\subsection{Nilpotent groups}\label{subsec:nilp}
In this subsection we consider nilpotent groups, which are usually considered as an immediate generalization of abelian groups.

\begin{definition}[Nilpotent groups]\label{def:nilp}
The \emph{lower central series} of a group $G$ is the inductively defined descending sequence of subgroups
\[
G = G_0 \geq G_1 \geq G_2 \geq \cdots,
\]
in which $G_k = [G, G_{k-1}]$.
A group $G$ is called \emph{nilpotent} if its lower central series terminates with $G_{d}$ being the trivial group for some finite $d$.
In this case, the smallest such $d$ is called the \emph{nilpotency class} of $G$.
\end{definition}

In particular, abelian groups are nilpotent of class one.
One can easily show $G_k \geq G^{(k)}$ by induction on $k$, therefore nilpotent groups are also solvable (but the converse is not true).
It is well known that all subgroups, quotients and direct products of nilpotent groups are nilpotent~\cite[Lemma~13.56, Theorem~13.57]{dructu2018geometric}.

\subsection*{The unitriangular matrix groups $\UT(n, \Q)$}

One of the most important examples of nilpotent groups is the group of unitriangular matrices:

\begin{definition}[Unitriangular matrix groups]\label{def:UT}
Denote by $\UT(n, \Q)$ the group of $n \times n$ upper triangular rational matrices with ones on the diagonal:
\[
\UT(n, \Q) \coloneqq 
\left\{
\begin{pmatrix}
1 & * & \cdots & * & * \\
0 & 1 & \cdots & * & * \\
\vdots & \vdots & \ddots & \vdots & \vdots \\
0 & 0 & \cdots & 1 & * \\
0 & 0 & \cdots & 0 & 1 \\
\end{pmatrix}
, \text{ where $*$ are entries in $\Q$ } \right\}.
\]
Then $\UT(n, \Q)$ is a nilpotent group of class $n-1$~\cite[Examples~13.36]{dructu2018geometric}.
Similarly, one can define the groups $\UT(n, \Z)$ by changing the entries from rationals to integers.
\end{definition}

The group $\UT(n, \Z)$ (and hence $\UT(n, \Q)$) play an important role in the study of finitely generated nilpotent groups by the following fact.
\begin{theorem}[{\cite{grunewald1980some}}]\label{fct:embed}
    Every finitely generated nilpotent group is isomorphic to a finite extension of a subgroup of $\UT(n,\Z)$ for some $n \in \N$.
\end{theorem}

We have already encountered the group $\HH_3(\Z) = \UT(3, \Z)$ in Subsection~\ref{subsec:dimthree}.
Therefore the results in this subsection can be seen as generalizations of Theorem~\ref{thm:heis}.

\begin{theorem}\label{thm:UT}
    In subgroups of $\UT(n, \Q)$, we have the following results.
    \begin{enumerate}[nosep, label = (\roman*)]
        \item \cite{malcev1953nilpotent} Group Membership is decidable in $\UT(n, \Q), n \in \N$.
        \item \cite{roman2022undecidability} There exists a large enough $k \in \N$, such that Semigroup Membership is undecidable in the direct product $\UT(3, \Z)^k$.
        \item \cite{DBLP:conf/stacs/000123} Let $G$ be a class-2 nilpotent subgroup of $\UT(n, \Q)$ for some $n$. Then Semigroup Intersection is PTIME decidable in $G$.
        \item \cite{dong2022identity} Fix $d \leq 10$. Let $G$ be a class-$d$ nilpotent subgroup of $\UT(n, \Q)$ for some $n$. Then the Identity Problem and the Group Problem are PTIME decidable in $G$.
    \end{enumerate}
\end{theorem}

For example, for any $k \in \N$, the direct product $\UT(3, \Z)^k$ is a class-two nilpotent subgroups of $\UT(3k, \Q)$.
Therefore Theorem~\ref{thm:nilp}(iii) shows that $\UT(3, \Z)^k$ admits decidable (PTIME) Semigroup Intersection.
However, for large enough $k$, it admits undecidable Semigroup Membership (Theorem~\ref{thm:UT}(ii)).

It is worth pointing out that Theorem~\ref{thm:UT}(i) follows from a more general result of Mal'cev on finitely generated nilpotent groups\footnote{While $\UT(n, \Q)$ is not finitely generated, for Group Membership we are always working in finitely generated subgroups of $\UT(n, \Q)$.} (see Theorem~\ref{thm:nilp}(i)).
Theorem~\ref{thm:UT}(ii) is proven through an embedding of the \emph{Hilbert's tenth problem}, whose undecidability follows from a celebrated result of Matiyasevich:

\begin{theorem}[{Hilbert's tenth problem~\cite{matiyasevich1970diophantineness}}]\label{thm:Hilbertten}
    The following problem is undecidable:
    given as input a polynomial $f \in \Z[X_1, \ldots, X_n]$, decide whether there exist integers $z_1, \ldots, z_n$ such that $f(z_1, \ldots, z_n) = 0$.
\end{theorem}

On the other hand, Theorem~\ref{thm:UT}(iii) and (iv) are proven using techniques from Lie algebra and convex geometry.
In fact, \cite{dong2022identity} proposed a conjecture (depending on the nilpotency class $d$), subject to which Theorem~\ref{thm:UT}(iv) still holds for $d > 10$.
The conjecture is only verified up to $d = 10$ using computer algebra software.

\subsection*{Arbitrary nilpotent groups}

The close connection between $\UT(n, \Q)$ and nilpotent groups allows one to generalize the decidability results in Theorem~\ref{thm:UT} to arbitrary finitely generated nilpotent groups.

\begin{theorem}\label{thm:nilp}
    In finitely generated nilpotent groups\footnote{As a convention in computational group theory, a finitely generated nilpotent group is usually represented by its \emph{finite presentation}, meaning it is written as a quotient $F(\Sigma) /\langle\langle v_1, \ldots, v_n\rangle\rangle$ for some alphabet $\Sigma$ and elements $v_1, \ldots, v_n \in \left(\Sigma^{\pm}\right)^*$, similar to the setup of the Word Problem. Every finitely generated nilpotent group admits a finite presentation~\cite[Proposition~13.84]{dructu2018geometric}.}, we have the following results.
    \begin{enumerate}[nosep, label = (\roman*)]
        \item \cite{malcev1953nilpotent} Group Membership is decidable in all finitely generated nilpotent groups.
        \item \cite{roman2022undecidability} There exists a class-2 nilpotent group where Semigroup Membership is undecidable.
        \item \cite{DBLP:conf/stacs/000123} Semigroup Intersection is decidable in all finitely generated class-2 nilpotent groups.
        \item \cite{dong2022identity} For $d \leq 10$, the Identity Problem and the Group Problem are decidable in all finitely generated class-$d$ nilpotent groups.
    \end{enumerate}
\end{theorem}

The decidability of the Identity Problem, the Group Problem and Semigroup Intersection in nilpotent groups of higher nilpotency classes remains open.

\begin{remark}\label{rmk:nilp}
    By the generalized Tits alternative~\cite{okninski1995generalised}, a subsemigroup of $\GL(n, \K)$ either contains a finite-index nilpotent subsemigroup\footnote{A semigroup is called nilpotent if the group it generates is nilpotent.} or it contains a subsemigroup isomorphic to the set of non-empty words $\{a, b\}^* \setminus \{\epsilon\}$.
    Similar to Remark~\ref{rmk:embedPCP}, Semigroup Intersection in the direct product $\big(\{a, b\}^* \setminus \{\epsilon\}\big) \times \big(\{a, b\}^* \setminus \{\epsilon\}\big)$ is undecidable due to the Post Correspondence Problem. 
    Therefore, nilpotent groups are the largest ``natural'' class of matrix groups (i.e.\ a class that is closed under direct products) where decidability of Semigroup Intersection remains possible.
\end{remark}

\subsection{Metabelian groups}\label{subsec:meta}

The class of metabelian groups is another generalization of abelian groups.

\begin{definition}[Metabelian groups]\label{def:meta}
A group $G$ is called \emph{metabelian} if it is solvable of derived length at most two.
Equivalently, $G$ is metabelian if and only if it contains an abelian normal subgroup $A$ such that the quotient $G/A$ is also abelian.
\end{definition}

All subgroups, quotients and direct products of metabelian groups are metabelian \cite[Proposition~13.91]{dructu2018geometric}.

\begin{example}
    Let $\T(2, \K)$ denote the group of $2 \times 2$ invertible upper-triangular matrices over any non-trivial field $\K$:
        \[
            \mathsf{T}(2, \K) \coloneqq 
            \left\{
            \begin{pmatrix}
            x & z \\
            0 & y \\
            \end{pmatrix}
            \;\middle|\; x, y \in \K \setminus \{0\}, z \in \K \right\}.
        \]
    Then $\T(2, \K)$ is metabelian but not nilpotent.
\end{example}

\subsection*{Example: the wreath product $\Z \wr \Z$}

An important example of metabelian groups is the \emph{wreath product} $\Z \wr \Z$.
The wreath product is a fundamental construction in group and semigroup theory, and it also plays an important role in the algebraic theory of automata.
The Krohn–Rhodes theorem~\cite{krohn1965algebraic} states that every finite semigroup (and correspondingly, every finite automaton) can be decomposed into elementary components using wreath products.

Following the work of \cite{magnus1939theorem}, it became clear that solving algorithmic problems in arbitrary metabelian groups usually boils down to solving them in wreath products of abelian groups~\cite{baumslag1973subgroups,baumslag1994algorithmic}.
Therefore, studying the group $\Z \wr \Z$ is usually the first step towards understanding general metabelian groups.

The wreath product $\Z \wr \Z$ is most easily defined as a matrix group over the Laurent polynomial ring $\Z[X, X^{-1}]$:
\begin{align}\label{eq:defphi}
    \Z \wr \Z = \left\{ 
\begin{pmatrix}
        X^{b} & y \\
        0 & 1
\end{pmatrix}
\;\middle|\; y \in \Z[X, X^{-1}], b \in \Z 
\right\}.
\end{align}
To see that $\Z \wr \Z$ is indeed metabelian, take its abelian normal subgroup 
\[
A \coloneqq \left\{ 
\begin{pmatrix}
        1 & y \\
        0 & 1
\end{pmatrix}
\;\middle|\; y \in \Z[X, X^{-1}] 
\right\} \cong \Z[X, X^{-1}],
\]
then the quotient $(\Z \wr \Z)/A \cong \Z$ is also abelian.

The group $\Z \wr \Z$ admits an interesting interpretation as an infinite-state machine.
Each element of $\Z \wr \Z$ can be seen as a configuration of a ``Turing machine''.
The machine is composed of a bi-infinite band with cells indexed by $\Z$, along with a \emph{head} positioned at one of its cells.
Each cell contains an integer.
The element $
\begin{pmatrix}
        X^{b} & \sum_{i = p}^q a_i X^i \\
        0 & 1
\end{pmatrix}
$ 
corresponds to a configuration where the cells contain the integers $\cdots, 0, a_{p}, a_{p+1}, \ldots, a_q, 0, \cdots$, with $a_i$ being on the $i$-th cell.
The head of the machine is placed on the $b$-th cell.
See Figure~\ref{fig:machine} for an illustration.

\begin{figure}[h]
    \centering
    \captionsetup{width=.9\linewidth}
    \medskip
    \begin{tikzpicture}[every node/.style={block},
            block/.style={minimum height=1.5em,outer sep=0pt,rectangle,node distance=0pt}]
            \node (mat) {$\begin{pmatrix}
                X^{b} & \cdots + a_{-1} X^{-1} + a_{0} + a_{1} X + \cdots \\
                0 & 1
                \end{pmatrix} \quad \Longleftrightarrow$};
           \node[draw] (0) [right=3.2cm of mat] {$a_0$};
           \node[draw] (-1) [left=of 0] {$a_{-1}$};
           \node[draw] (-2) [left=of -1] {$a_{-2}$};
           \node (-3) [left=of -2] {$\cdots$};
           \node[draw] (1) [right=of 0] {$a_{1}$};
           \node[draw] (2) [right=of 1] {$a_{2}$};
           \node (3) [right=of 2] {$\cdots$};
           \draw[->] (0.south)+(0, -4mm) -- (0.south);
           \draw[->] (2.south) -- ++(0, -4mm);
           \draw (-2.north west) -- ++(-1cm,0) (-2.south west) -- ++ (-1cm,0) 
                         (2.north east) -- ++(1cm,0) (2.south east) -- ++ (1cm,0);
    \end{tikzpicture}
    \medskip
    \caption{Intepretation of an element in $\Z \wr \Z$ as a ``Turing machine''. The upward arrow $\uparrow$ marks the $0$-th cell, and the downward arrow $\downarrow$ marks the position of the head. In this example, $\downarrow$ marks the cell at position 2, meaning $b = 2$.}
    \label{fig:machine}
\end{figure}

\begin{figure}[h]
    \centering
    \captionsetup{width=.9\linewidth}
    \begin{tikzpicture}[every node/.style={block},
            block/.style={minimum height=1.5em,outer sep=0pt,rectangle,node distance=0pt}]
           \node[draw] (0) {$\red{1}$};
           \node[draw] (1) [right=of 0] {$\red{2}$};
           \node[draw] (15) [right=of 1] {$\red{2}$};
           \draw[->] (0.south)+(0, -4mm) -- (0.south);
           \draw[->, lightgray] (1.south) -- ++(0, -4mm);

           \node (mat1) [above=0.6cm of 1] {\;$\begin{pmatrix}
                X^{1} & 1 + 2X + 2X^2 \\
                0 & 1
                \end{pmatrix} \quad\;\; \times$};

           \node (times) [right= 1.2cm of 15] {$\times$};

           \node[draw] (2) [right= 1.2cm of times] {$\blue{3}$};
           \node[draw] (3) [right= of 2] {$\blue{3}$};
           \node[draw] (4) [right= of 3] {$\blue{3}$};
           \draw[->, lightgray] (2.south)+(0, -4mm) -- (2.south);
           \draw[->] (4.south) -- ++(0, -4mm);

           \node (mat2) [above=0.6cm of 3] {$\quad\;\; \begin{pmatrix}
                X^{2} & 3 + 3X + 3X^2 \\
                0 & 1
                \end{pmatrix} \;\; = \; $};

           \node (eq) [right= 1.2cm of 4] {$=$};

           \node[draw] (5) [right= 1.2cm of eq] {$\red{1}$};
           \node[draw] (6) [right= of 5] {$\purple{5}$};
           \node[draw] (7) [right= of 6] {$\purple{5}$};
           \node[draw] (8) [right= of 7] {$\blue{3}$};
           \draw[->] (5.south)+(0, -4mm) -- (5.south);
           \draw[->] (8.south) -- ++(0, -4mm);

           \node (mat3) [above=0.6cm of 7] {\; $\begin{pmatrix}
                X^{3} & 1 + 5X + 5X^2 + 3X^3 \\
                0 & 1
                \end{pmatrix}$};
    \end{tikzpicture}
    \medskip
    \caption{Example of multiplying two elements in $\Z \wr \Z$ using the matrix representation and using the machine representation.}
    \label{fig:grouplaw}
\end{figure}

Multiplication in $\Z \wr \Z$ also admits a simple interpretation using this machine representation.
To multiple two elements under the machine representation, it suffices to align the $\downarrow$ arrow of first element and the $\uparrow$ arrow of second element, then add up the integers in corresponding cells.
For the arrows, we keep the position of $\uparrow$ in the first element and the position of $\downarrow$ in the second element.
See Figure~\ref{fig:grouplaw} for an illustration.

Using the machine representation, every element in $\Z \wr \Z$ can be seen as an instruction on the machine by its effect of right-multiplication.
For example, the element
$\begin{pmatrix}
    X & 0 \\
    0 & 1
\end{pmatrix}$ or
$
\begin{tikzpicture}[baseline=-1.6ex, every node/.style={block},
    block/.style={minimum height=1.5em,outer sep=0pt,rectangle,node distance=0pt}]
    \node[draw] (0) {$0$};
    \node[draw] (1) [right=of 0] {$0$};
    \draw[->] (0.south)+(0, -4mm) -- (0.south);
    \draw[->] (1.south) -- ++(0, -4mm);
\end{tikzpicture}$
corresponds to the instruction ``move the head one cell to the right'', which is the effect of (right)-multiplying 
$
\begin{tikzpicture}[baseline=-1.6ex, every node/.style={block},
    block/.style={minimum height=1.5em,outer sep=0pt,rectangle,node distance=0pt}]
    \node[draw] (0) {$0$};
    \node[draw] (1) [right=of 0] {$0$};
    \draw[->] (0.south)+(0, -4mm) -- (0.south);
    \draw[->] (1.south) -- ++(0, -4mm);
\end{tikzpicture}$ to a configuration.
Similarly, the element
$\begin{pmatrix}
    1 & 1 \\
    0 & 1
\end{pmatrix}$ or
$
\begin{tikzpicture}[baseline=-1.6ex, every node/.style={block},
    block/.style={minimum height=1.5em,outer sep=0pt,rectangle,node distance=0pt}]
    \node[draw] (0) {$1$};
    \draw[->] (0.south)+(0, -4mm) -- (0.south);
    \draw[->] (0.south) -- ++(0, -4mm);
\end{tikzpicture}$
corresponds to the instruction ``in the cell marked by the head, increase the number by one''.
Hence, Semigroup Membership in $\Z \wr \Z$ can be reformulated as the problem: given a configuration $T$ and a set of instructions $\mG$, can one reach $T$ using a sequence of instructions from $\mG$?
Similarly, the Identity Problem can be reformulated as: given a set of instructions $\mG$, can one reach the initial configuration using a (non-empty) sequence of instructions from $\mG$?
For $\Z \wr \Z$, we have the following results.

\begin{theorem}\label{thm:ZwrZ}
    In the group $\Z \wr \Z$:
    \begin{enumerate}[nosep, label = (\roman*)]
        \item \cite{lohrey2015rational} Semigroup Membership is undecidable.
        \item \cite{romanovskii1974some} Group Membership in decidable.
        \item \cite{DBLP:conf/icalp/000123c} The Identity Problem and the Group Problem are decidable.
    \end{enumerate}
\end{theorem}

Given the machine-like structure of $\Z \wr \Z$, the undecidability result in Theorem~\ref{thm:ZwrZ}(i) may not come as a surprise.
Indeed, \cite{lohrey2015rational} showed that Semigroup Membership in $\Z \wr \Z$ can simulate the halting problem in \emph{2-counter machines} (also known as \emph{Minsky machines}), which is equivalent to the halting problem in arbitrary Turing machines~\cite{minsky1967computation}.
As for Theorem~\ref{thm:ZwrZ}(ii), it follows from a more general result of Romanovskii on metabelian groups (see Theorem~\ref{thm:meta}(ii)).
The most unexpected result seems to be the decidability of the Identity Problem in Theorem~\ref{thm:ZwrZ}(iii).
This shows that by restricting the target to the neutral element in Semigroup Membership, the problem becomes much more tractable (the same phenomenon can be observed in Theorem~\ref{thm:nilp}).
The proof of Theorem~\ref{thm:ZwrZ}(iii) uses a combination of graph theory techniques and deep tools from algebraic geometry.

\begin{rmk}\label{rmk:ZwrZ}
    The decidability of Semigroup Intersection in $\Z \wr \Z$ remains an open problem at the moment.
    It can be shown that $\Z \wr \Z$ does not contain $\{a, b\}^* \times \{a, b\}^*$ as a subsemigroup.
    This excludes the possibility of proving undecidability by directly embedding the Post Correspondence Problem.
    However, $\Z \wr \Z$ contains $\{a, b\}^*$ as a subsemigroup~\cite{baumslag1982groups}.
    Therefore as in Remark~\ref{rmk:embedPCP}, Semigroup Intersection is undecidable in the direct product $(\Z \wr \Z) \times (\Z \wr \Z)$.
\end{rmk}

\subsection*{Arbitrary metabelian groups and beyond}

We now consider arbitrary finitely generated metabelian groups.

\begin{theorem}\label{thm:meta}
    In finitely generated metabelian groups\footnote{As a convention in computational group theory, a finitely generated metabelian group is usually represented by a \emph{finite $\mathscr{A}^2$-presentation}.
    Let $\Sigma$ be an alphabet and $F(\Sigma)$ be the free group over $\Sigma$.
    The quotient $M(\Sigma) \coloneqq F(\Sigma)/[[F(\Sigma), F(\Sigma)], [F(\Sigma), F(\Sigma)]]$ is metabelian, and is called the \emph{free metabelian group} over $\Sigma$.
    A finite $\mathscr{A}^2$-presentation of a metabelian group $G$ is the writing of $G$ as a quotient $M(\Sigma)/\langle \langle v_1, \ldots, v_n \rangle \rangle$ for some alphabet $\Sigma$ as well as elements $v_1, \ldots, v_n \in M(\Sigma)$.
    Here, $\langle \langle v_1, \ldots, v_n \rangle \rangle$ denotes the normal subgroup of $M(\Sigma)$ generated by $v_1, \ldots, v_n$.
    Every finitely generated metabelian group admits a finite $\mathscr{A}^2$-presentation~\cite[Corollary~1]{hall1954finiteness}~\cite[p.629]{baumslag1994algorithmic}.}, we have the following results.
    \begin{enumerate}[nosep, label = (\roman*)]
        \item \cite{lohrey2015rational} There exists a finitely generated metabelian group, namely $\Z \wr \Z$, where Semigroup Membership is undecidable.
        \item \cite{romanovskii1974some} Group Membership is decidable in all finitely generated metabelian groups.
        \item \cite{dong2023semigroup} The Identity Problem and the Group Problem are decidable in all finitely generated metabelian groups.
        \item (See Remark~\ref{rmk:ZwrZ}) There exists a finitely generated metabelian group, namely $(\Z \wr \Z) \times (\Z \wr \Z)$, where Semigroup Intersection is undecidable.
    \end{enumerate}
\end{theorem}

In particular, Theorem~\ref{thm:meta}(iii) generalizes Theorem~\ref{thm:ZwrZ}(iii), while Theorem~\ref{thm:ZwrZ}(ii) is in fact a corollary of Theorem~\ref{thm:meta}(ii).

Beyond nilpotent groups and metabelian groups, it would be interesting to consider algorithmic problems in larger classes of solvable groups.
A natural candidate is the class of \emph{polycyclic groups}~\cite[Chapter~13, p.463]{dructu2018geometric}, which generalizes finitely generated nilpotent groups.
By a classic result of~\cite{kopytov1968solvability}, polycyclic groups admit decidable Group Membership.
However, they are not yet well-studied in the context of semigroup algorithmic problems.
Another interesting class to consider is the so-called \emph{center-by-metabelian groups}~\cite{groves1978finitely}, which are a relatively tractable subclass of solvable groups of derived length three.

\section{Other algorithmic problems for groups and semigroups}\label{sec:other}
To end this survey we mention some other interesting algorithmic problems in infinite groups and matrix groups.
These problems are less directly related to semigroup theory, but have nevertheless tight connections with many other areas in the theory of computing.
Let $G$ be a group. The following list is non-exhaustive.

\begin{enumerate}[noitemsep, label = (\roman*)]
    \item (Knapsack Problem~\cite{myasnikov2015knapsack,konig2016knapsack}) Given as input the elements $A_1, \ldots, A_K, T \in G$, decide whether there exist $n_1, \ldots, n_K \in \N$ such that $A_1^{n_1} \cdots A_K^{n_K} = T$.
    \item (Freeness Problem~\cite{cassaigne1999undecidability,cassaigne2012decidability}) Given as input the elements $A_1, \ldots, A_K \in G$, decide whether there exist two different sequences $(i_1, \ldots, i_p)$ and $(j_1, \ldots, j_q)$ such that $A_{i_1} \cdots A_{i_p} = A_{j_1} \cdots A_{j_q}$.
    \item (Subgroup Intersection~\cite{howson1954intersection,baumslag2010subgroups}) Given two finite sets $\mG, \mH \subseteq G$, decide whether $\gmG \cap \langle \mH \rangle_{grp}$ is the trivial group.
    \item (Coset Intersection~\cite{babai2010coset,macdonald2019low}) Given two finite sets $\mG, \mH \subseteq G$, as well as elements $g, h \in G$, decide whether $g \cdot \gmG \cap h \cdot \langle \mH \rangle_{grp}$ is empty.
    \item (Vector Reachability~\cite{bell2008reachability,potapov2019vector}) Given $K$ square matrices $A_1, \ldots, A_K$ of dimension $d$, as well as two vectors $v, w$ of dimension $d$, decide whether there exists $T \in \langle A_1, \ldots, A_K \rangle$ such that $Tv = w$.
    In particular, if one only considers invertible matrices, then Vector Reachability can be reformulated as a generalization of Coset Intersection.
    Let $T_0$ be an arbitrary matrix such that $T_0 v = w$, and denote by $H$ the matrix group $\{A \mid Aw = w\}$.
    Then Vector Reachability is equivalent to deciding whether the semigroup $\langle A_1, \ldots, A_K \rangle$ has non-empty intersection with the coset $H \cdot T_0$.
    \item (Skolem Problem~\cite{skolem1933einige,ouaknine2015linear}) Given a square matrix $A$ of dimension $d$, as well as two vectors $v, w$ of dimension $d$, decide whether there exists $n \in \N$ such that $v^{\top} A^n w = 0$.
    \item (Positivity Problem~\cite{soittola1975dol,ouaknine2014positivity}) Given a square matrix $A$ of dimension $d$, as well as two vectors $v, w$ of dimension $d$, decide whether there exists $n \in \N$ such that $v^{\top} A^n w \geq 0$.
\end{enumerate}


\bibliographystyle{ACM-Reference-Format-Journals}
\bibliography{advances}

\end{document}